\documentclass[%
 aip,
 sd,%
 amsmath,amssymb,
preprint,%
]{revtex4-1}

\usepackage{graphicx}
\usepackage{dcolumn}
\usepackage{bm}

\draft 

\begin{document}


\title{A partially wetting film of water and surfactant under the influence of a propagating MHz surface acoustic wave} 



\author{Gennady Altshuler}
\author{Ofer Manor}
 \email{manoro@technion.ac.il}
 \homepage[]{http://lab-sst.net.technion.ac.il}
\affiliation{Wolfson Department of Chemical Engineering, Technion---Israel Institute of Technology, Haifa 32000, Israel}



\date{\today}

\begin{abstract}
We use both theory and experiment to study the response of {\it partially wetting} films of water and surfactant solutions to a propagating MHz vibration in the solid substrate in the form of a Rayleigh surface acoustic wave (SAW). The SAW invokes a drift of mass in the liquid film, which is associated with the Schlichting boundary layer flow (also known as the Schlichting streaming). We study thin films that are governed by a balance between the drift and capillary stress alone. We demonstrate weak capillary contributions, such as for silicon oil films, support dynamic wetting and lead to the spreading of the liquid over the solid substrate along the path of the SAW. Strong capillary contributions, such as for water films, support however a concurrent dynamic wetting and dewetting along the path of the SAW, such that the film displace along the solid substrate. In addition, such films may support the formation of a capillary train-wave that propagate along the the same path. We further note the mechanism for film dynamics we discuss here is different to the more familiar Eckart streaming mechanism, which is associated with a film thickness that is greater than the wavelength of the sound leakage off the SAW and usually observed to support the motion of drops. The thickness of the films we discuss here is small in respect to the wavelength of the sound leakage, rendering contributions from the Eckart streaming, acoustic radiation pressure, and the attenuation of the SAW small.    
\end{abstract}


\maketitle 


\section{Introduction}
In this paper we use both theory and experiment to investigate the wetting and de-wetting dynamics of partially wetting films of water and surfactant solutions under the influence of a MHz propagating vibration in the form of a Rayleigh surface acoustic wave (SAW) in the solid substrate. The dynamics of films of liquid and the transfer of momentum from a mechanical vibration in a solid substrate, such as surface and bulk acoustic waves (SAWs, BAWs), or in a fluid, such as sound waves or surface waves, to flow has been separately investigated for many years now. For example, films of liquid have been studied for actuating microfluidic platforms \citep{Atencia:2005bu,Stone:2004kg,Whitesides:2006jj}, cooling electronic platforms \citep{Amon:2001ji,BarCohen:2006if}, desalinating sea water \citep{Fletcher:1974jz}, manufacturing \citep{Fendler:1996gb,Nagayama:1996vd,Wang:2004kx}, etc., and the transfer of momentum from a vibration or a wave to flow, generally known as acoustic streaming, has been employed to study underwater sea currents \citep{oceanphysics}, sound waves near obstacles \citep{TheoreticalAcoustics}, SAW microfluidics \citep{yeoannurev}, etc. 

Both fields of study were recently converged in order to explore the dynamic wetting of liquid films under the influence of a propagating MHz SAW. Previous studies explored {\em fully wetting} (at equilibrium) films of silicon oil\citep{Rezk:2012dx,rezkprsa,PRE2015} and {\em partially wetting} (at equilibrium) films of water and surfactant solutions\citep{GennadyPoF2015}. In the second mentioned study we explored films that are thick enough to support the accumulation of sound waves that diffract (leak) off the SAW in the substrate. Here we compliment this previous study and investigate the dynamics of films of {\em partially wetting} solutions of water and a surfactant that are too thin to support the accumulation of the diffracting sound waves. We show the response of the thin {\em partially wetting} films to the SAW qualitatively differs from the response of the thicker films and from the response of {\em fully wetting} films of similar thickness in previous studies. 

The primary mechanism that support film dynamics in the above mentioned studies on film dynamics and in the present study is a directional drift of liquid, invoked by the SAW \citep{Manor:2012be}. This mechanism is associated with the Schlichting boundary layer flow (also known as the Schlichting streaming) \citep{Stokes,Rayleigh:1884p254,Schlichting:1932p447}. This mechanism is different to the the Eckart streaming \citep{Eckart:1948to}, the widely employed mechanism for introducing drop motion by exciting a SAW in the solid substrate. The Eckart streaming mechanism appears in the presence of a SAW due to sound waves that diffract (leak) off the SAW and into the liquid \citep{Arzt:1967p112,Shiokawa:1989tg}. An appreciable viscous attenuation of the sound wave leakage in the liquid supports the formation of a secondary vortical flow field, i.e., the Eckart streaming, which is capable of displacing liquid drops \citep{Brunet:2010p668}. Here however we study films of liquid that are too thin to support the sound waves that diffract off the SAW, such that Eckart streaming is avoided. The dominating drift of mass in such films appears due to the direct contact between the component of the SAW at the solid surface and the neighbouring liquid.

Briefly, a SAW in the substrate excites a periodic viscous flow in the neighbouring liquid within the viscous penetration length $\delta\equiv\sqrt{2\mu/\rho\omega}$, where $\mu,~\rho$ and $\omega$ are the viscosity and density of the liquid and the vibration angular frequency, respectively. Additional convective contributions further support a drift of liquid mass, also known as the Stokes drift \citep{Stokes} or the Rayleigh streaming \citep{LIGHTHILL:1978p12}. A propagating SAW support a directional drift along the path of the SAW \citep{Manor:2012be}. 

%

The SAW diffracts (leaks) same frequency sound waves of wavelength $\lambda$ into the liquid film \citep{Campbell:1970wb,Shiokawa:1989tg}, imposing a steady convective stress at its free surface, which is known as the acoustic radiation pressure \citep{King:1934tp,Anonymous:5ILrl7tg,Hasegawa:2000vy}. In a previous study \citep{GennadyPoF2015} we considered the dynamics of liquid films with a thickness $h$ that was comparable in size to the wavelength of the sound leakage, $h\approx\lambda$. Eckart streaming was weak and acoustic radiation pressure, drift of liquid mass, and capillary stress dominated the dynamic of the liquid films. The films were observed to spread opposite the path of the propagating SAW when convective contributions from the drift dominated the capillary stress. In this study, however, we generate films that are thin in respect to the wavelength of the diffracted sound waves, $h\ll\lambda$, so that the sound waves do not accumulate in the films. The attenuation of the SAW under the films is weak and ignored, and the film is governed by a balance between the drift of liquid mass due to the SAW and the capillary stress \citep{GennadyPoF2015}. 


\begin{figure}
  \includegraphics[height=6cm]{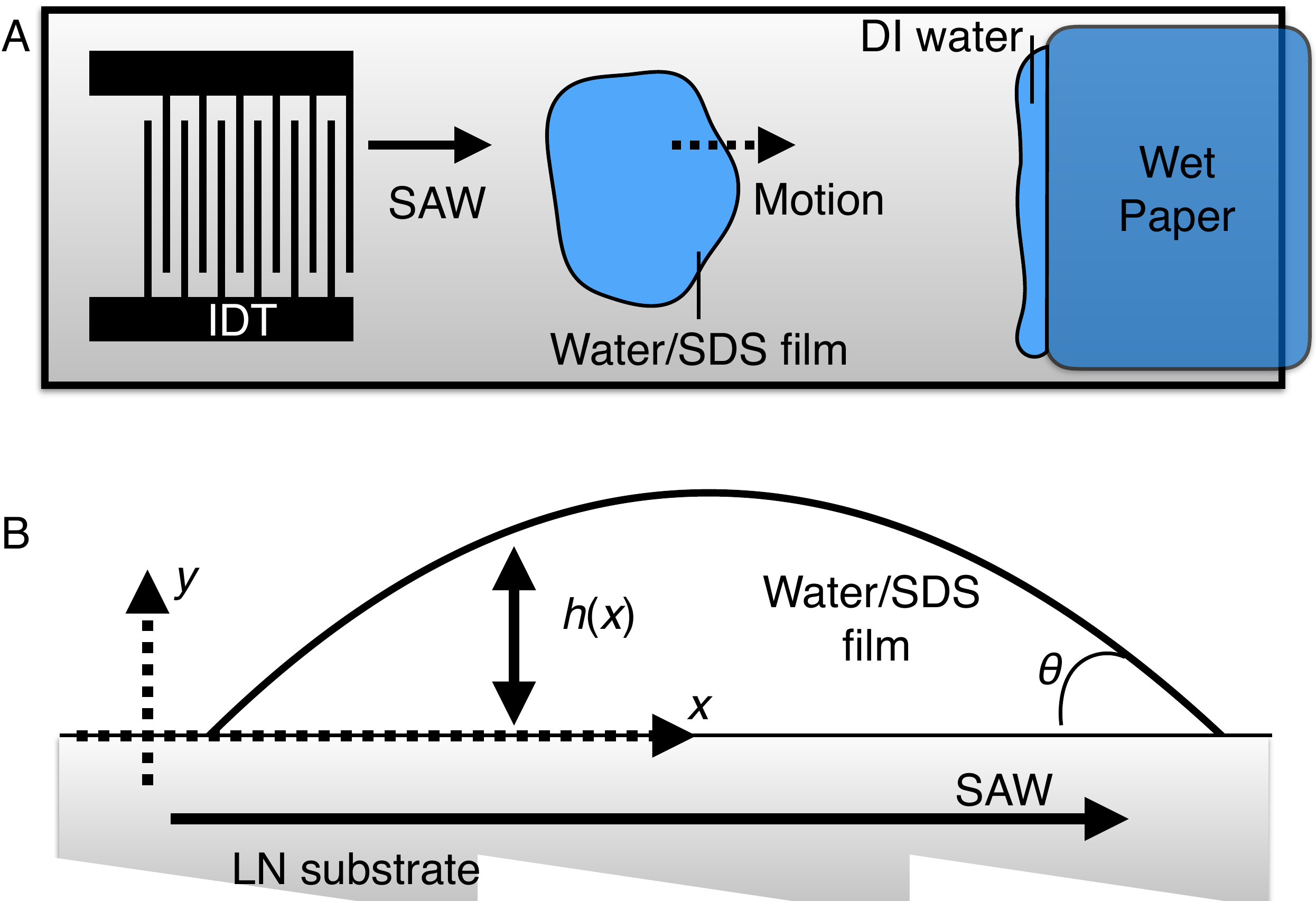}
\caption{Illustrations of (A) the upper view and (B) the side view of the water/SDS film (the contact angle is small, $\theta < 30^\circ$) atop a SAW device, comprising a piezoelectric lithium niobate substrate, patterned with metallic interdigitated electrodes that produce a propagating Rayleigh SAW in the substrate to be eventually absorbed by a thick paper wick that contains de-ionised (DI) water for eliminating SAW reflections off the edges of the device.}
  \label{system}
\end{figure}

\begin{figure*}
 \centering
 \includegraphics[height=10cm]{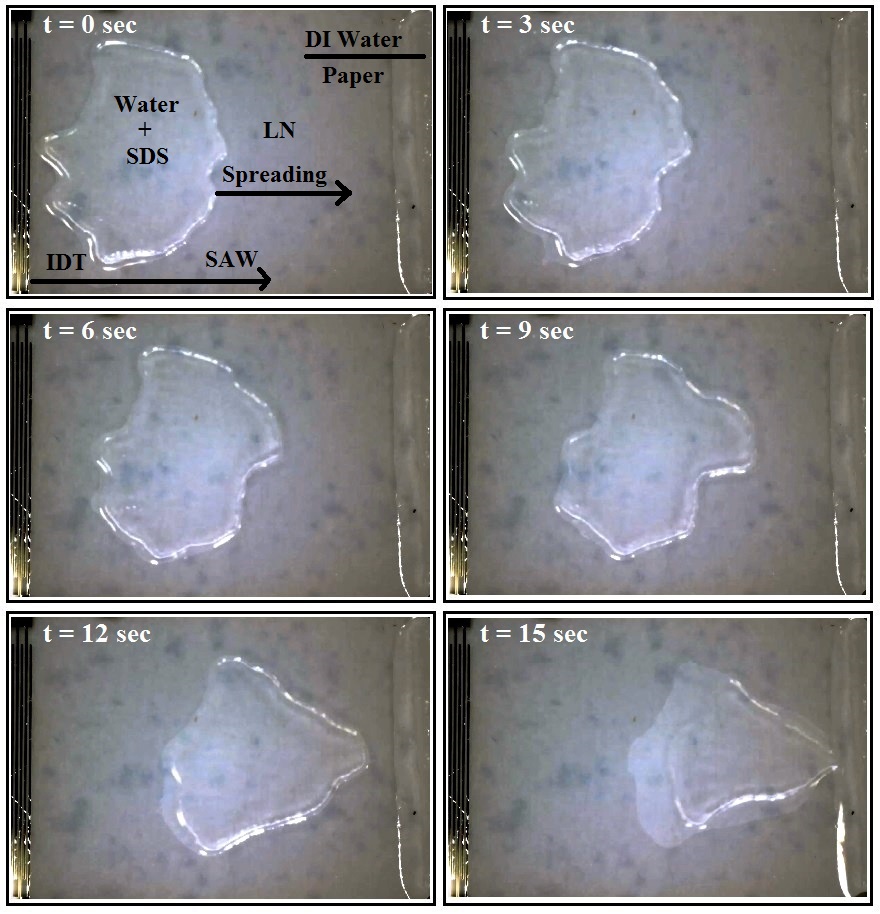}
\caption{A series of images portraying the dynamics of the water/SDS film, illustrated in figure \ref{system}, under the influence of a 20 MHz Rayleigh SAW that propagates in the lithium niobate (LN) substrate. The water film spread from left to right along the path of the SAW, which is emitted from the interdigitated electrodes to the left and absorbed by the wet paper wick to the right of the device. The IDT electrodes in the image are 50 micron wide and are separated from each other by a  distance of 50 micron, which illustrates the length scale in the images.}
 \label{expshow}
\end{figure*}

In this paper we use theory to study the physics of the dynamic wetting and de-wetting of a {\it partially wetting} liquid film in contact with a propagating MHz Rayleigh SAW in the solid substrate, illustrated in figure \ref{system}A, and demonstrate our theoretical insights in laboratory, shown in figure \ref{expshow}. We explore the physical mechanisms that govern the film dynamics in \S \ref{Theory}, where we discuss the governing equations in \S \ref{Equations} and simulate the dynamics of the film in \S \ref{Simulation}. We further compare theory and experiment in \S \ref{Experiment} and demonstrate the response of thin films of water and a surfactant solutions to a MHz SAW. We conclude our findings in \S \ref{Conclusions}.


\section{Theory}\label{Theory} 
\subsection{Equations}\label{Equations}
We consider a liquid film of thickness $h$ that possesses a finite three phase contact angle $\theta$ with the solid surface at equilibrium. We then assume the film is excited by a Rayleigh SAW in the solid substrate, illustrated in figure \ref{system}B. The SAW, propagating in the solid along the $x$ coordinate, support both an in-plane ($x_m$) and an out-of-plane ($y_m$) displacements of a similar magnitude at the solid surface. The leading order components of the surface motion are \citep{elasticwaves} 
	 \begin{equation}
	     \left( \begin{array}{c} x_m ,~ y_m \end{array} \right)_{y=0}  =
	      \left( \begin{array}{c}  \displaystyle  x+(U/\omega) x_1 ,~  \displaystyle (U/\omega)y_1 \end{array} \right),
	      \label{DisplacementSAWDIM}
	    \end{equation}    
where $x_1=\sin(\omega t-k x)$, $y_1=\chi\cos(\omega t-k x)$, and $U,~\chi,\omega,~k$, and $t$ are the velocity amplitude of the solid surface, ratio between the velocity amplitude of the in-plane and out-of-plane displacements of the solid surface ($\chi\approx1.3$ in an isotropic solid), SAW angular frequency, SAW wavenumber, and time, respectively. The expression in (\ref{DisplacementSAWDIM}) is taken from the linear theory for Rayleigh waves, giving $x_m$ and $y_m$ at the undeformed plane of the solid surface, where $y=0$. This result may be taken to be the leading order approximation for the motion components of the Rayleigh wave at the solid surface for $U/\omega\ll k^{-1}$, where the influence of the solid deformation on the Rayleigh wave is neglected. The leading order surface velocity associated with the surface displacement in (\ref{DisplacementSAWDIM}) is then
	 \begin{equation}
	     \left( \begin{array}{c} \dot{x_m} ,~ \dot{y_m} \end{array} \right)_{y=0} =
	      \left( \begin{array}{c}  \displaystyle  U \dot{x_1} ,~  \displaystyle U\dot{y_1} \end{array} \right).
	      \label{VelocitySAWDIM}
	    \end{equation}    
The tangent and normal stresses at the free surface of the liquid film are satisfied by
\begin{equation}  
	      \left( \begin{array}{c} \partial_y u_x \\ \partial_y u_y \end{array} \right)_{y=h} =
	      \left( \begin{array}{c}   0 \\ (2\mu)^{-1} (2\gamma \kappa+p-\Pi) \end{array} \right),
	      \label{bcSurfaceDIM}
	    \end{equation}    
where $\gamma \kappa$ is the capillary stress in which $\kappa\approx \partial_{xx} h/2$ is the mean curvature of the free surface of the film, $\mu$ and $\gamma$ are the liquid viscosity and the liquid/vapour surface tension, respectively, $p$ is an excess hydrostatic pressure in the film and $-\Pi$ is an excess disjoining pressure over the atmospheric pressure above the film that accounts for intermolecular forces. 
 In addition, the thickness of the film $h$ is governed by the kinematic condition:
 \begin{equation}
 \partial_th+u_x\partial_xh =u_y, \quad \text{at}~y=h.
 \label{kinematicDIM}
  \end{equation} 
    
Momentum and mass conservations in the film, the above conditions, and assuming $k^{-1}\gg\delta\gg U/\omega$ (which is generally appropriate when exciting water at ambient conditions using MHz SAWs) give the film equation \citep{PRE2015,GennadyPoF2015}
	\begin{equation}
	 \partial_t h=\displaystyle -{\rm Re} U \partial_x\left(hf\left({h}/{\delta}\right)\right)-\frac{1}{3\mu}\partial_x \left(h^3\left( \gamma\partial_{x}^3h-\partial_x\Pi\right)\right).
	    \label{film1}
	\end{equation}
		where ${\rm Re}\equiv \rho U \delta/\mu$ and 
	\begin{eqnarray*}
&f(\zeta)=\displaystyle -\frac{1}{\zeta}\frac{\cos (2\zeta)+\cosh (2\zeta)-6 \cos (\zeta) \cosh (\zeta)+4}{4 (\cos (2\zeta)+\cosh (2\zeta)-8 \cos (\zeta) \cosh (\zeta)+8)}+\\ \nonumber
&\displaystyle \frac{-\sin (2\zeta)+\sinh (2\zeta)-4 \cos (\zeta) \sinh (\zeta)+4 \sin (\zeta) \cosh (\zeta)}{4 (\cos (2\zeta)+\cosh (2\zeta)-8 \cos
   (\zeta) \cosh (\zeta)+8)}-\\ \nonumber
&\displaystyle \zeta\frac{\sin (\zeta) \sinh (\zeta)}{2 (\cos (2\zeta)+\cosh (2 \zeta)-8 \cos (\zeta) \cosh (\zeta)+8)}
\label{f}
\end{eqnarray*}
is the non-dimensional average velocity (scaled by ${\rm Re}U$) of the liquid in the film along the solid surface (along the $x$ coordinate), where $\zeta\equiv{h}/{\delta}$. Large film thickness, such that $h\gg\delta$, renders $f \to 1/4$ and small film thickness, such that $h\ll\delta$, gives $f \to 0$. 


We use the transformations
 \begin{equation} 
  \label{scale2}
 h\to H~h,~x\to (H/\theta\epsilon)~x,~(p,~\Pi) \to \left(\gamma\theta^2\epsilon^2/H\right)(p,~\Pi)
  \end{equation}  
 to render the film equation non-dimensional, where $H$ is the characteristic thickness of the film and $\epsilon$ mediates between the two characteristic length scale in this problem $\delta$ and $H$. The viscous penetration length $\delta$ governs the drift velocity that scales like ${\rm Re}U$. The characteristic film thickness $H$ governs the magnitude of the capillary stress in the film that scales like $\gamma/H$. 
The choice $\epsilon\equiv (\delta/H)^{1/3}$ eliminates $\epsilon$ from the non-dimensional film equations given below \citep{GennadyPoF2015}.  
 
 
 
  The response of the film to the SAW may be associated with two characteristic velocities and corresponding time scales. If contributions from the SAW govern the film then the velocity in the film scales with ${\rm Re} U$ -- the characteristic velocity of the drift of liquid mass. The corresponding transformation for the time is then $t\to \left(H/{\rm Re}U\theta\epsilon\right) t$, giving the non-dimensional film equation 
	\begin{equation}
	\partial_t h=\displaystyle -\partial_x\left(hf\left(\frac{H}{\delta}h\right)\right)-\frac{\theta^3}{3\rm{We}}\partial_x \left(h^3\left( \partial_{x}^3h-\partial_x\Pi\right)\right).
	    \label{film2}
	\end{equation}
If however contributions from the capillary stress govern the dynamics of the film then the velocity in the film scales with $\epsilon^3\theta^3\gamma/\mu$ -- the augmented capillary velocity of the film. The corresponding transformation for the time is then $t\to \left(\mu H/(\epsilon\theta)^4\gamma\right) t$, giving the non-dimensional film equation  
	\begin{equation}
	 \partial_t h=\displaystyle -\left(\frac{\theta^3}{\rm{We}}\right)^{-1}\partial_x\left(hf\left(\frac{H}{\delta}h\right)\right)-\frac{1}{3}\partial_x \left(h^3\left( \partial_{x}^3h-\partial_x\Pi\right)\right). 
	 	    \label{film3}
	\end{equation}
Both the film equations in (\ref{film2}) and (\ref{film3}) are augmented kinematic wave equations such that in both cases the film dynamics is along the path of the propagating SAW. Both equations are governed by the non-dimensional parameter $\theta^3/{\rm We}$, where ${\rm We}\equiv \rho U^2 H/\gamma$. This parameter measures contributions from the capillary stress and the drift of mass. Further detail is given elsewhere \citep{GennadyPoF2015}. 

The film equation in \ref{film2} is associated with ${\theta^3}/{\rm{We}}<1$, suggesting the film is dominated by contributions from the SAW. The film equation in \ref{film3} is associated with ${\theta^3}/{\rm{We}}>1$, suggesting the film is dominated by the capillary stress. In the following analysis we show {\em fully wetting} films of silicon oil, studied elsewhere \citep{rezkprsa,PRE2015}, are associated with ${\theta^3}/{\rm{We}}\ll1$. The {\em partially wetting} films of water and a surfactant solutions, we study here, are however associated with ${\theta^3}/{\rm{We}}>1$. The geometry of the {\em partially wetting} films is governed by the capillary stress while the drift of liquid in the films renders motion.


\subsection{Simulation}\label{Simulation}
Here we simulate the dynamics of the liquid film by solving the film equations in (\ref{film2}) and (\ref{film3}). We assume the macroscopic liquid film possesses an initial parabolic geometry. In dimensional terms this translates to $h={\text max}\left(h_p+(1-x^2)(H-h_p),~h_p\right)$ at $t=0$, where $H$ is the maximal thickness of the initial film geometry and $h_p$ is the thickness of a thin precursor film, wetting the solid surface ahead of the macroscopic body of liquid and elevating viscous stresses near the three phase contact line. The precursor film is assumed flat and at rest far away from the drop, giving $\partial_xh=\partial_{xxx}h=0$ at $x\to\pm\infty$. For ease of the numerical analysis we set the precursor film thickness $h_p$ at 10\% of $H$.

 Further, the three phase contact angle $\theta$ is governed by the dimensional disjoining pressure $\Pi=3\gamma\theta^2h_p^{-1}\left((h_p/h)^3-(h_p/h)^n\right)$ that vanishes at $h=h_p$ and $h\gg h_p$. The first term in $\Pi$ is associated with a repulsive van der Waals force between the solid and free surface of the liquid film. The second term is associated with an attractive shorter range force that may be associated with a secondary contribution of the van der Waals force\citep{Pismen:2002uj}, for which n=6, or the Lennard Jones potential, for which n=9. Another option is to approximate the contribution of an electrical double layer force for which the second term should be exponential \citep{Teletzke:1988ei}. We employ\citep{Schwartz:1998uv} $n=4$ for ease of the numerical procedure since the macroscopic properties of the film are known to be a weak function of the attractive term in the disjoining pressure as long as it decays faster than the van der Waals force \citep{Schwartz:1998uv}, i.e., $n>3$. We integrate the numerical system using a second order finite difference algorithm in space, employing 1000 to 4000 nodes, and an alternating direction implicit algorithm in time, using up to 10,000 steps, to demonstrate film dynamics at different values of the non-dimensional parameter $\theta^3/{\rm We}$.   

\begin{figure}[h]
\centering
  \includegraphics[height=10cm]{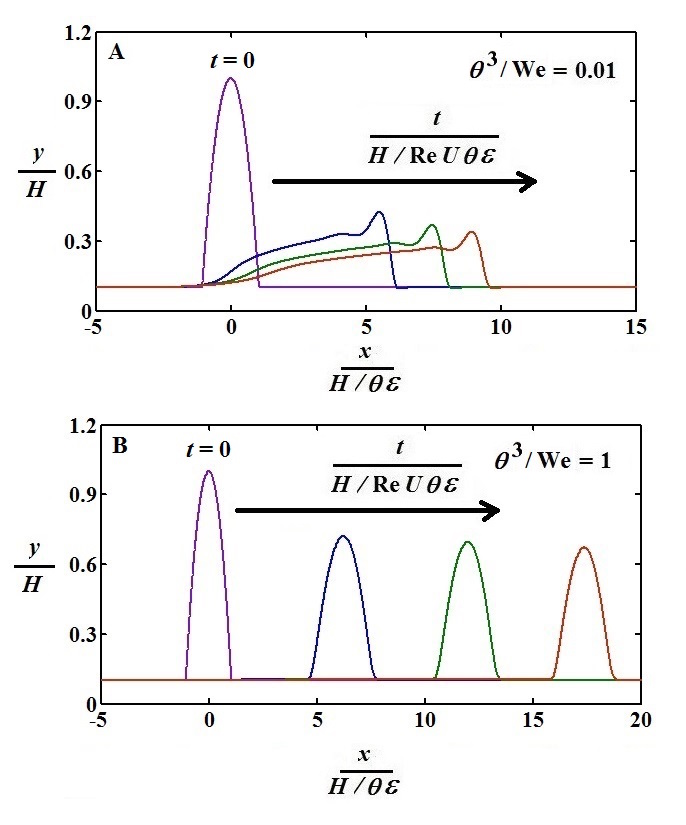}
\caption{Time variations of the film geometry should (A) contributions from the convective drift of mass invoked by the SAW dominate the film dynamics and (B) contributions from both the capillary stress and the drift of mass dominate the film dynamics, where the SAW propagates in the substrate from left to right.}
\label{evopic}
\end{figure}

\begin{figure}[h]
\centering
  \includegraphics[height=10cm]{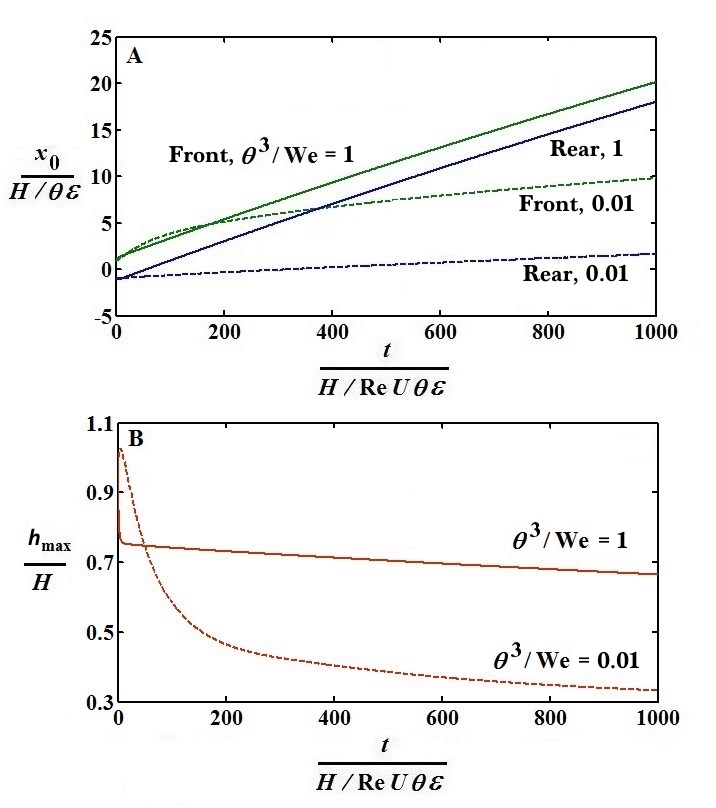}
\caption{Time variations of (A) the position of the rear (receding contact line) and the front (advancing contact line) of the liquid film and (B) the maximal thickness of the film should contributes from the drift of mass dominate the film dynamics (dashed) and should contributions from both the capillary stress and the drift of mass dominate the film dynamics (solid).}
\label{paramstudy}
\end{figure}

When the drift of mass dominates the film dynamics the film satisfies $\theta^3/{\rm We}\ll1$. The drift of mass mainly supports the motion of the front of the film (advancing contact line). The rear of the film (receding contact line) moves at a similar direction at a much slower pace, shown in figures \ref{evopic}A and \ref{paramstudy}A for $\theta^3/{\rm We}=10^{-2}$. The drift vanishes at the front and rear of the liquid film, where the thickness of the film becomes small, since $f(H/\delta)\ll1$ for $h/\delta\ll1$. However, the film equation in (\ref{film2}) supports the accumulation of liquid mass near the front (a kinematic wave dynamics) due to the drift. This increases the local curvature near the front, elevating the magnitude of the capillary stress necessary for increasing film thickness in the absence of the drift. The front of the film advances then along the path of the SAW.

Further, the drift transport mass from the rear of the film to its front. The decrease in the thickness of the film at the rear decreases the local curvature of the free surface of the film and thus the local capillary stress. Thus the velocity at which the rear of the film de-wet the solid surface is slow in respect to the velocity the front of the film wet the solid surface. Since contributions from the capillary stress to film dynamics are generally small the free surface of the film continuously increases in area, rendering a weak global connection between the front and the rear of the film. This is illustrated in the continuous variation of the maximal thickness of the film in figure \ref{paramstudy}B. Thus, the weak local and global contributions to the rear of the film inhibit the motion of the rear of the film. Similar results were observed in laboratory in previous studies where low surface tension and {\em fully wetting} (in equilibrium) silicon oil films were excited by SAWs \citep{Rezk:2012dx,rezkprsa}.

When the capillary stress and the drift are balanced the film dynamics satisfies $\theta^3/{\rm We}\approx1$. The film dynamics in figure \ref{evopic}B is then qualitatively different. we observe an initial stage where the film spread isotropically towards (although not achieving) a capillary equilibrium. This is shown in the fast initial reduction of the film maximal thickness from $h_{\text{max}}=1$ to $h_{\text{max}}\approx0.75$ in figure \ref{paramstudy}B, very close to the time the film is deposited atop the solid surface (at $t=0$). Then, near capillary equilibrium the film responds to the convective drift of mass and both the front (advancing contact line) and the rear (receding contact line) of the liquid film move at similar rates along the path of the propagating SAW. The rates are globally related due to the capillary constraint on the area of the free surface of the film that roughly conserves the film geometry, following the initial capillary spreading of the film. This is shown in figure \ref{paramstudy}B in terms of the small variations of the maximal thickness of the film with time (following the initial and rapid stage of capillary spreading). A dominating capillary stress will then follow this dynamics, further restricting the geometry of the film.


\begin{figure}
 \centering
 \includegraphics[height=14cm]{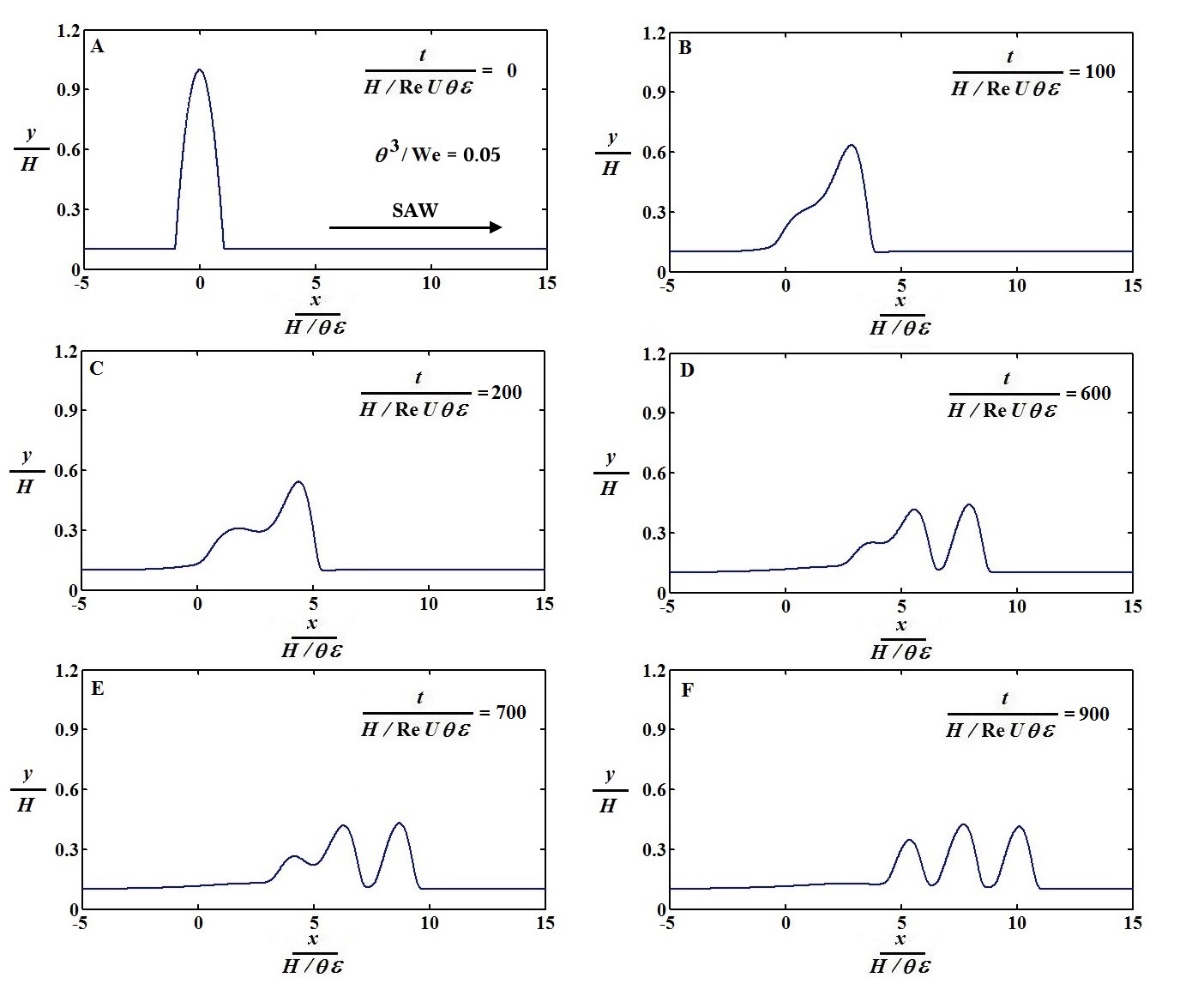}
\caption{Spatial variations of the film geometry at different times for $\theta^3/{\rm We}=0.05$ where the SAW propagates in the substrate from left to right and the film appears to evolve into three consecutive waves that propagate along the same path.}
\label{extreme1}
\end{figure}

\begin{figure}
 \centering
 \includegraphics[height=14cm]{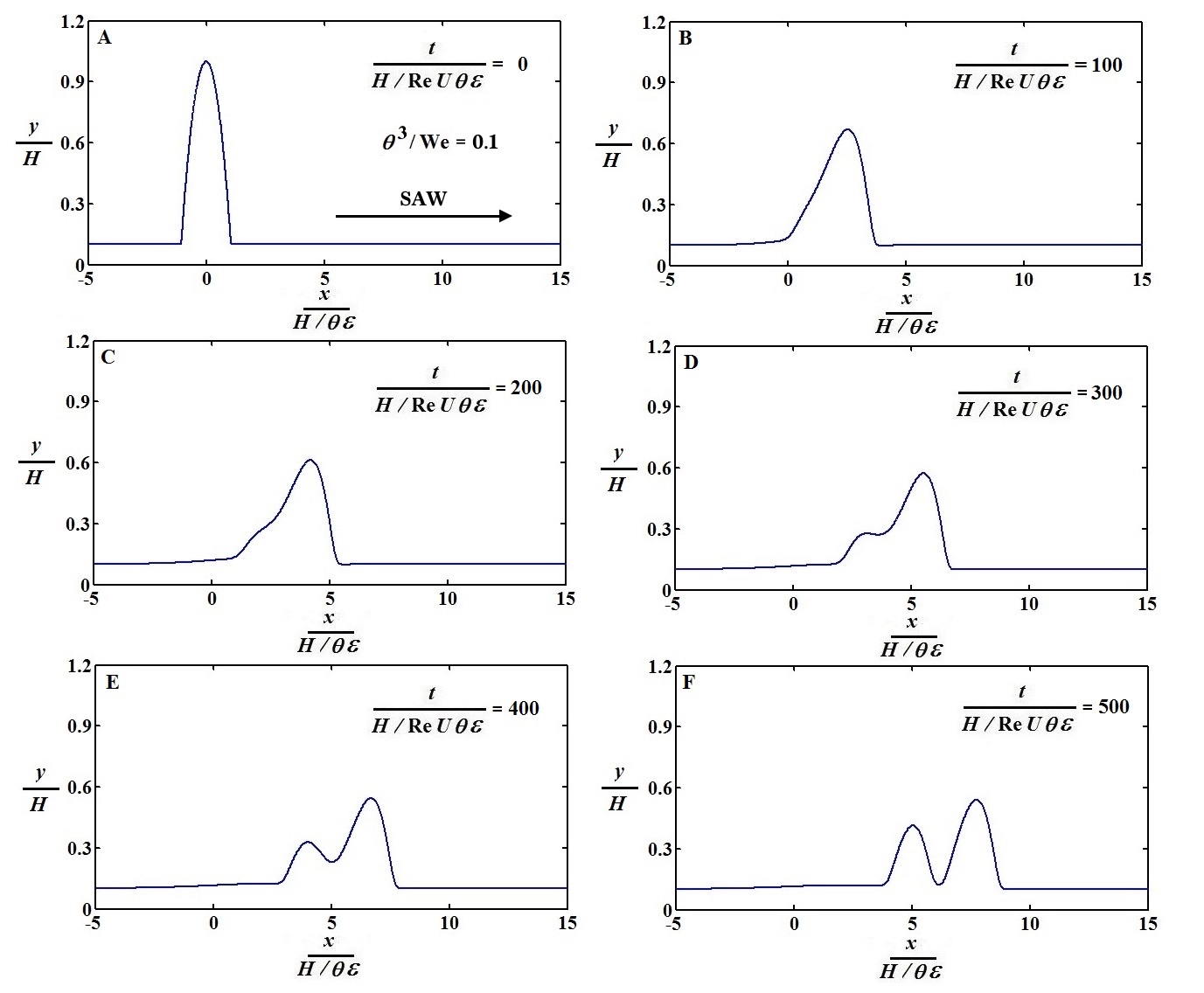}
\caption{Spatial variations of the film geometry at different times for $\theta^3/{\rm We}=0.1$ where the SAW propagates in the substrate from left to right and the film appears to evolve into two consecutive waves that propagate along the same path.}
\label{extreme2}
\end{figure}

Both the capillary and convective (SAW induced drift of mass) mechanisms contribute to the dynamic variations of the film geometry for $\theta^3/{\rm We}$ values between the two limits we discuss above. This interplay of mechanisms may further give the non trivial dynamics that we show in figures \ref{extreme1} and \ref{extreme2} for $\theta^3/{\rm We}=0.05$ and 0.1, respectively. More specifically, under the right balance between both mechanisms the film may break to a wave train that propagates along the path of the SAW. This result was observed earlier in laboratory \citep{Comment1}. This phenomenon is different in nature to a previous report on the formation of wave trains off much thicker films of silicon oil, where the motion of the wave trains was opposite to the spreading direction of the liquid film and was attributed to contributions from acoustic radiation pressure and Eckart streaming \citep{Rezk:2012dx}.

 In the following section we use a laboratory procedure to study the dynamics of {\em partially wetting} films of water and surfactant under the influence of a MHz SAW, where the capillary stress dominates the film dynamics. Here we complement a previous study, where we excited thin films of silicon oil using SAWs that were governed by the same mechanisms \citep{rezkprsa}. We use the theory to scale the raw experimental data, which many times appears without specific trend. The scaled data collapse to attain specific trends along with variations in the non dimensional number employed in our theory, suggesting the mechanisms given in our theory govern the experiment. We avoid quantitative comparison between theory and experiment due to the use of a qualitative model for the precursor film in our theory.


\section{Experiment}\label{Experiment}

We now use experiment to explore the response of thin films of water and the surfactant sodium dodecyl sulfate (SDS) to a MHz Rayleigh SAW that propagate in the solid substrate. We generate a Rayleigh SAW by employing 20 MHz SAW devices that are comprised of a 10 nm chrome / 100 nm aluminium interdigital transducer (IDT), patterned using a standard photolithography, on a 0.5 mm thick $128^\circ$ Y--cut, X--propagating, single crystal lithium niobate piezoelectric substrate, illustrated in figure \ref{system}. The SAW was generated by applying a sinusoidal electric voltage to the IDT using a signal generator (SMB100A, Rohde\&Schwartz) and an amplifier (BSA 5001-2, Bonn Elektronik). In order to avoid SAW reflections we absorbed the SAW near the edges of the devices using an absorber. Our wave absorber was a paper wick (Cleanroom Wipe, Texwipe) wetted with DI water, which we found to be insensitive to the SAW excitation \citep{GennadyPoF2015}. to wet the paper wick We further used a Laser Doppler Vibrometer (MSA 500, Polytec) to measure the frequency and velocity amplitude of the Rayleigh SAW at the solid surface and to verify it is a purely propagating SAW.

We prepared partially wetting liquids by adding small amounts of the surfactant sodium dodecyl sulfate (SDS) (Specially Pure S.G., Sigma Aldrich) to de-ionised (DI) water up to the critical micelle concentration (CMC). The three phase contact angle $\theta$ of the water/SDS solutions was measured using a goniometer (model 250, Rame-Hart) against different surfactant concentrations; the values of $\theta$ we use are associated with the advancing three phase contact angle. Further detail is given elsewhere \citep{GennadyPoF2015}. Values of the corresponding surface tension of the water and water and SDS solutions were taken from the literature \citep{Dean:2007it}. We captured images of the liquid films using a USB microscope (model AD7013MZ, Dino-Lite) and further used MATLAB for image analysis. 

Prior to measurement we cleaned the surface of the SAW devices by dipping in acetone, isopropanol, and DI water and subsequently drying using nitrogen. We then used a thin brush to deposit strips of the liquid films atop the solid substrate near the IDT. We excited the films using a propagating Rayleigh SAW in the solid substrate, shown in figure \ref{system}. The films were observed to spread and displace along the path of the SAW while slowly loosing mass due to evaporation, shown in figure \ref{expshow}. 

\begin{figure}[h]
\centering
  \includegraphics[height=15.5cm]{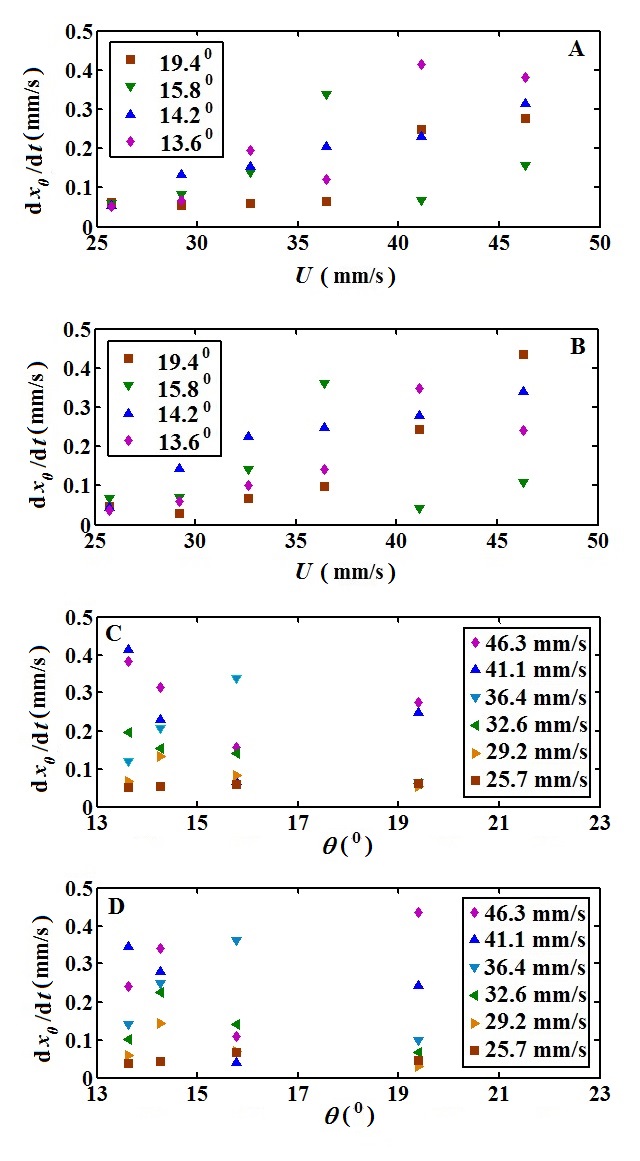}
\caption{SAW intensity ($U$) variations of the velocities of the (A) rear and (B) front of the water and surfactant films for different values of the contact angle $\theta$, and contact angle ($\theta$) variations of the (C) rear and (D) front of the films for different values of the SAW intensity $U$.}
\label{uthin}
\end{figure}

In figure \ref{uthin} we give the dimensional velocities of the front (advancing contact line) and the rear (receding contact line) of the film for different contact angles, $\theta$, and for different SAW intensities (velocity amplitudes of the solid surface), $U$. It appears the velocities of both the front and the rear of the film increase with increasing $U$, although there is no clear relation to the contact angle of the liquid film in the measured contact angle range $\theta=13^\circ-20^\circ$ (associated with a concentration of $8-2$ mM SDS in water, respectively). 

Films of pure DI water that give $\theta=28^\circ$ did not appear to respond to the SAW in the parameter range we employed in our measurements. Further, some of the water and SDS films, we deposited on the SAW device, did not appear to initially respond to the SAW. The films commenced motion after several seconds of evaporation. 

In order to perceive the absence or delayed film response to the SAW we recall a previous analysis \citep{rezkprsa,PRE2015} where we showed that films may spread along the path of the propagating SAW should $h<\lambda/8$. Otherwise contributions from sound waves that diffract off the SAW to the attenuation of the SAW and acoustic radiation pressure may become appreciable and alter the dynamics of the film. The analysis accounted for local contributions from the capillary stress, the drift of mass, and sound waves in the film that may support acoustic radiation pressure, akin to low surface tension silicon oil films. Considering however the high surface tension of the films in this study we should further expect an additional integral contribution to this rule from the overall constraint over the free surface of the films by the capillary stress, highlighted in \S \ref{Simulation}. To first approximation however we admit a possible reason for the absence of a film response to the SAW is that its thickness is greater than $\lambda/8$. 

Further, the wavelength of the 20 MHz SAW we employed in the solid substrate was $2\pi k^{-1}\approx200~\mu$m. The wavelength of the diffracted sound wave in the liquid film was then $\lambda\approx75~\mu$m due to the approximate ratio of 0.375 between the phase velocity of the sound in water and of the Rayleigh SAW in the crystal cut of the lithium niobate substrate we used. Approximating the film geometry in figure \ref{system}B as a parabola we may approximate the maximal initial film thickness to be $h_{\text{max}}\approx L\tan{|\theta|}/2$, where $L$ is the characteristic width of the liquid film. The contact angle range $\theta\approx13^\circ-20^\circ$ gives $h_{\text{max}}\approx 9-18~\mu$m for $L\approx1$ mm. We then observe the initial maximal thickness of the film $h_{\text{max}}$ is likely to be close to the film thickness threshold $\lambda/8\approx 9~\mu$m for dynamic wetting; this may explain the tendency of some of the films to commence spreading after several seconds of evaporation. Further, the contact angle of DI water $\theta\approx28^\circ$ gives  $h_{\text{max}}\approx 27~\mu$m, which is three times the threshold of the film thickness, suggesting a greater resistance to motion than in water/SDS films.

\begin{figure}[h]
\centering
  \includegraphics[height=15.5cm]{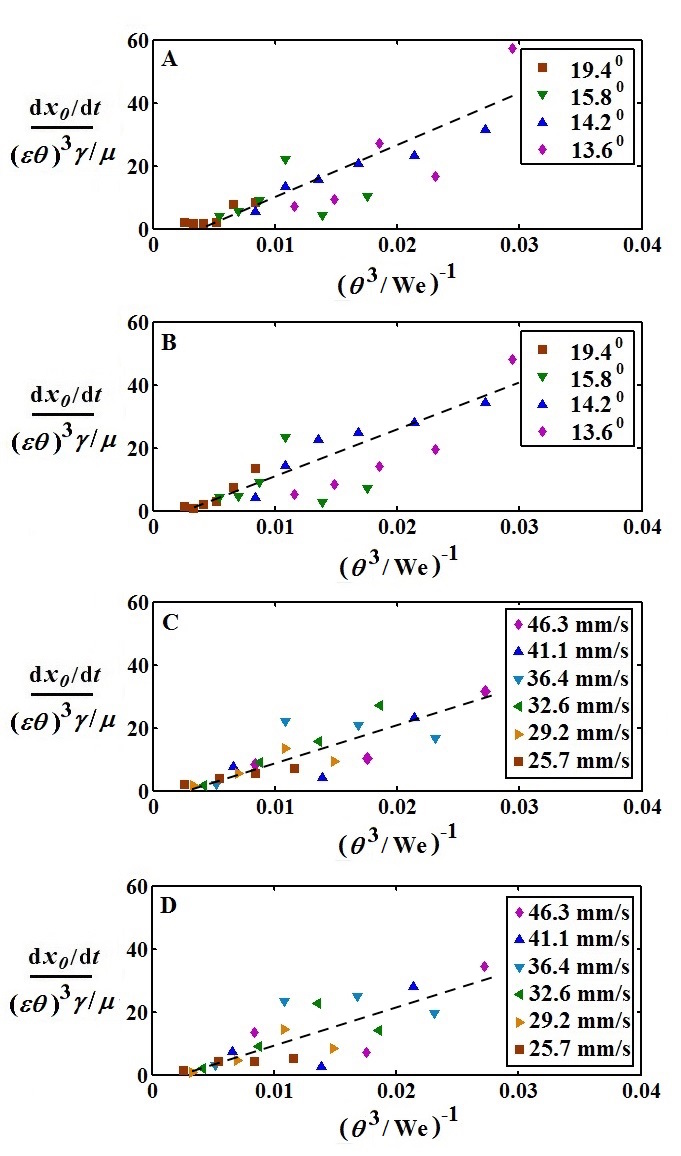}
\caption{Parametric ($\theta^3/{\rm We}$) variations of the normalised velocity of the (A) rear and (B) front of the water and surfactant films for different values of the contact angle $\theta$ and the (C) rear and (D) front of the films for different values of the SAW intensity $U$. The dashed lines illustrate the general trend of data.}
\label{uthinn}
\end{figure}

\begin{figure}[h]
\centering
  \includegraphics[height=4cm]{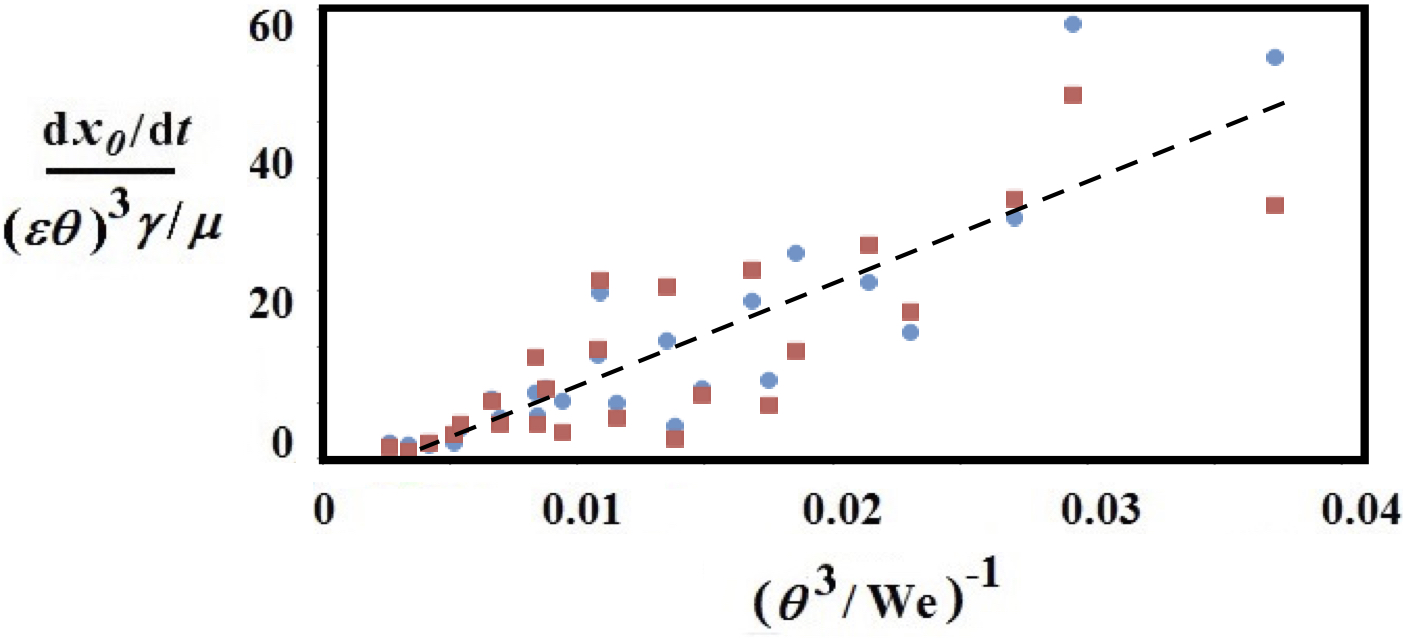}
\caption{Parametric ($\theta^3/{\rm We}$) variations of the normalised velocities of both the rear (blue circles) and front (red squares) of the water and surfactant films. The dashed line illustrates the general trend of data.}
\label{condat}
\end{figure}

We define the characteristic thickness $H\equiv\lambda/8\approx 9~\mu$m for the following analysis of our experimental results. Our experiments with water and surfactant solutions give $\theta^3/{\rm We}>1$. The film dynamics corresponds then to the film equation in (\ref{film3}). Thus, in figure \ref{uthinn} we normalise the velocities of the front and the rear of the film by $\epsilon^3\theta^3\gamma/\mu\equiv (\delta/H)(\theta^3\gamma/\mu)$ and plot the result against the non-dimensional parameter $(\theta^3/{\rm We})^{-1}$. We observe a monotonous increase in the normalised velocity with a decrease in $\theta^3/{\rm We}$. We further depicted both the velocities of the rear and the front of the films in figure \ref{condat}, suggesting there is no clear rule that differentiates the velocities of both ends of the films in the parameter range we employ in the experiment. This result is in agreement with the theoretical result in figure \ref{paramstudy}, where for $\theta^3/{\rm We}=1$ the velocities of the front and the rear of the film are similar due to the dominating capillary mechanism that constrains the area of the free surface of the film.


We further note there are additional mechanisms that are likely to influence the film dynamics. The most prominent of which is the above mentioned evaporative loss of water volume. A different possible mechanism that may contribute to the film dynamics is the distribution of surface tension, also known as the Marangoni mechanism, due to a non even evaporative heat loss or due to the convection of the surfactant. Another possible mechanism is an increase in the film viscosity and a decrease in the surface tension due to an increase in the concentration of the surfactant to result from evaporation. However, the monotonous trend of the data in figures \ref{uthinn} and \ref{condat} suggests that while these additional mechanisms may add to the distribution (noise) of data, the main contribution to the dynamics of the film is associated with the balance between the convective contribution of the propagating Rayleigh SAW, i.e., the drift of mass, and the capillary stress, given in (\ref{film3}). 

The result in figure \ref{condat} further strengthen our notion that contributions from the evaporation and Marangoni mechanisms to the velocity of the front and the rear of the film are small. Loss of water mass due to evaporation should support an inward motion of both the front and the rear of the film towards its interior, which is the case for the evaporation of a liquid film in the absence of the SAW. In the presence of the SAW, where the film is in motion due to the drift of mass, this mechanism should increase the velocity of the rear of the film and decrease the velocity of the front of the film in a laboratory frame of view. Further, the presence of the Marangoni mechanism, which is the result of convection of heat or the surfactant, should impose asymmetry of stress at the free surface between the front and the rear of the film. The results depicted in Fig. \ref{condat} state however there is no clear rule to differentiate between the velocities of the rear and the front of the film that appear to follow our theoretical insights.

\section{Conclusion}\label{Conclusions}
In this paper we study the dynamic wetting of {\em partially wetting} (at equilibrium) films of water and surfactant solutions under the influence of a propagating MHz Rayleigh SAW. The film thickness is small in respect to the wavelength of the sound wave in the liquid that diffract (leak) off the SAW in the solid, so that the sound does not accumulate within the liquid film. Mechanisms such as the acoustic radiation pressure, the Eckart streaming, and the attenuation of the SAW are weak and ignored. 

The dominating flow in the film is viscous and periodic to result from the direct contact between the liquid and the solid surface, which is set into a harmonic motion by the propagating SAW. This flow further supports a convective drift of liquid mass along the path of the Rayleigh SAW, and is usually known as the Schlichting boundary layer flow or the Schlichting streaming. Contributions to film dynamics from the drift of mass and from the capillary stress may support dynamic wetting and de-wetting. 

The ratio between the contributions of the drift of mass and the capillary stress to the film dynamics is given in the non-dimensional parameter $\theta^3/{\rm We}$. We show that {\em fully wetting} (at equilibrium) silicon oil films, where the drift of mass dominates the film, satisfy $\theta^3/{\rm We}\ll1$. The velocity of the front of the film scales with the order of magnitude of the drift velocity ${\rm Re}U$ and advance along the path of the propagating SAW. However, the rear of the films appears still or moves at a slower pace, such that the free surface of the film continuously increases in area. Intermediate contributions of the capillary stress and the drift of liquid mass to film dynamics may further break the liquid film to wave trains.

This is different to {\em partially wetting} (at equilibrium) films of water and surfactant solutions that experience an appreciable capillary stress in addition to the drift of liquid mass, such that  $\theta^3/{\rm We}>1$. The rear and the front of the film are then connected by an appreciable capillary stress, constraining the area of the free surface of the film. Motion of the front of the film renders then a similar motion at the rear of the film. The experimental velocities of the front and the rear scale then with the modified capillary velocity ${(\theta\epsilon)^3\gamma/\mu}\equiv {(\delta/H)(\theta^3\gamma/\mu)}$, which is in agreement with our theory. 

Further, the normalised experimental results give a monotonous trend along with variations in the reciprocal of the non-dimensional parameter $\theta^3/{\rm We}$. The collapse of the scaled experimental data suggest we have captured the appropriate physics in our theory. Additional contributions to the film dynamics such as the evaporation of water and the convection of surfactants are then small in the parameter range we employed in our experiments.


%
%

%

\begin{acknowledgments}
We acknowledge funding in support of this work by the Israel Science Foundation, grant number 489/14.
\end{acknowledgments}

\bibliography{references} 
\bibliographystyle{rsc} 

\end{document}